\documentclass[twocolumn,showpacs,preprintnumbers,amsmath,amssymb]{revtex4}
\usepackage{graphicx}% Include figure files
\usepackage{dcolumn}% Align table columns on decimal point
\usepackage{bm}% bold math

\begin{document}
\title{Magnetization-induced optical second-harmonic generation and local surface
plasmons in magnetic Co$_{x}$Ag$_{1-x}$ nanogranular films}

\author{E.M. Kim$^{1}$}
\author{T.V. Murzina$^{1}$}
\author{A.F. Kravets$^{2}$}
\author{M. Inoue$^{3}$}
\author{G. Marowsky$^{4}$}
\author{O.A. Aktsipetrov$^{1\ast}$}

\affiliation {$^{1}$Department of Physics, Moscow State University,
Moscow, 119992 Russia\\ $^{2}$Institute of Magnetism, National
Academy of Sciences of Ukraine, Kiev, 03680 Ukraine\\
$^{3}$Toyohashi University of Technology, Toyohashi 441-8580 Japan\\
$^{4}$Laser-Laboratorium G\"ottingen, D-37077 G\"ottingen, Germany}

\date{\today}% It is always \today, today,

\begin{abstract}
{Spectroscopy of magnetization-induced second-harmonic generation
(MSHG) is studied in magnetic Co$_{x}$Ag$_{1-x}$ granular films
containing Co nanoparticles. A strong resonance of the magnetic
contrast of the second-harmonic generation (SHG) intensity is
observed in the two-photon energy range from 3,8 eV to 4,5 eV. The
local surface plasmons exited in magnetic Co nanogranules in this
spectral range assist MSHG and are responsible for a significant
enhancement of the SHG magnetic contrast due to the resonances of
local optical fields.}
\end{abstract}

%\pacs{78.67.-n,42.65.Ky, 42.70.Qs}% PACS, the Physics and Astronomy
\maketitle

Linear magneto-optics with its more than a century-long history,
remains one of the most important experimental methods in studies of
magnetism. Meanwhile, significant attention has been recently
directed towards the development of nonlinear magneto-optics [1]:
magnetization-induced second-harmonic generation (MSHG) was observed
experimentally in yttrium-iron-garnet films [2], surfaces of
magnetic metals [3], in magnetic multilayers [4] and nanogranules
[5]. Experimental measurements [2-5] and theoretical estimates [6]
reveal the typical magnitudes of the magnetization-induced effects
in second-harmonic generation (SHG): the magnetization-induced
variations of the SHG intensity and rotation of the second-harmonic
(SH) wave polarization, may exceed the linear magneto-optical Kerr
effect (MOKE) by orders of magnitude. The term nonlinear
magneto-optical Kerr effect (NOMOKE) is usually used for MSHG in the
reflection geometry. As a result of these studies it was recognized
that MSHG is a powerful probe of nanomagnetism.

One of the important classes of magnetic nanostructures are magnetic
nanogranular films, e.g. Co$_{x}$Ag$_{1-x}$ films. Below the
percolation threshold ($x<0.45$) Co$_{x}$Ag$_{1-x}$ films are arrays
of Co nanoparticles embedded into nonmagnetic Ag matrix. These
Co$_{x}$Ag$_{1-x}$ granular alloys exhibit giant magnetoresistance
effect (GMR) [7].

Apart from extraordinary magneto-transport properties, arrays of
magnetic metal nanoparticles are expected to exhibit unusual optical
effects, e.g. plasmon assisted linear MOKE is considered in magnetic
nanoparticles [8,9]. In this relation, one of the astonishing
optical effects in metal nanoparticles is surface-enhanced SHG that
was observed by Wokaun, et al. [10] in silver island films. The
enhancement of the SHG intensity by up to three orders of magnitude
was attributed in Ref. [10] to the resonant enhancement of the local
field at the SH wavelength, mediated by the excitation of the local
surface plasmons (LSPs) in silver nanoparticles. This plasmon
mechanism of the SHG enhancement is profoundly studied for the last
decades (see [11] and references in this paper). According to
phenomenological approach, second-order nonlinear polarization of an
array of small metal particles is given by:
$\textbf{P}(2\omega)=L_{\alpha}^{2\omega} \hat{\chi}^{(2)}(2\omega)
\vdots L_{\alpha}^{\omega} \textbf{E}(\omega) L_{\alpha}^{\omega}
\textbf{E}(\omega)$, where $\hat{\chi}^{(2)}(2\omega)$ is the
second-order susceptibility of metal; $\textbf{E}(\omega)$ is the
optical field at fundamental wavelength; $L_{\alpha}^{\omega}$ and
$L_{\alpha}^{2\omega}$ are the anisotropic local field (LF) factors
at the fundamental and SH wavelengths, respectively, and the symbol
$(\vdots)$ relates to the convolution of the nonlinear
susceptibility tensor and vectors of fundamental optical field.

The spectral dependence of the anisotropic LF factor of an array of
small metal spheroids embedded in a dielectric matrix, within the
dipole and effective media approximations, is given by [12]:

\begin{equation}
\displaystyle
L_{\alpha}(\lambda)=\frac{\varepsilon_{d}(\lambda)}{\varepsilon_{d}(\lambda)
+[\varepsilon_{m}(\lambda)-\varepsilon_{d}(\lambda)](N_{\alpha}-\beta
x)}, \label{1}
\end{equation}
where $\varepsilon_{d}(\lambda)$ and $\varepsilon_{m}(\lambda)$ are
the complex dielectric constants of the dielectric matrix  and of
the metal at the wavelength $\lambda$, respectively; $\beta$ is
Lorenz field factor and $x$ is the filling factor, i.e. the relative
fraction of the metal in a composite material; $N_{\alpha}$ is an
anisotropic shape-dependent depolarization factor of the spheroids;
subscript $\alpha=\parallel,\perp$ denotes the tangential and normal
orientation of principle semiaxes $a$ and $b$ of the spheroids with
respect to the sample surface ( see the left-hand inset in Figure
1b).

\begin{table*}
    \begin{tabular}{|l|l|c|}
      \hline
      & Nonmagnetic susceptibility,  $\chi_{ijk}^{(2)even}$
      & Magnetic susceptibility,  $\chi_{ijk}^{(2)odd}(M||Y)$ \\

      \hline
      & $\chi^{(2)cryst}_{\perp zz}, \chi^{(2)cryst}_{\perp xx}=\chi^{(2)cryst}_{\perp yy}, \chi^{(2)cryst}_{\parallel xz}=\chi^{(2)}_{\parallel yz}$

      & $\chi^{(2)}_{\parallel yy}(M), \chi^{(2)}_{\parallel xx}(M),\,\chi^{(2)}_{\parallel yx}(M),
      \chi^{(2)}_{\parallel zz}(M),\,\chi^{(2)}_{\perp zx}(M)$\\

      \hline

\end{tabular}
\caption{Elements of nonmagnetic $\chi_{ijk}^{(2)even}$ and magnetic
$\chi_{ijk}^{(2)odd}(M||Y)$ susceptibility tensor and pseudotensor;
the latter are selected for transversal configuration of nonlinear
magneto-optical Kerr effect, i.e. M $\parallel$ Y (see the top inset
of Figure 1a). The first subscripts associated with the nonlinear
polarization at 2$\omega$ are denoted as $\parallel$ and $\perp$ to
be attributed to corresponding components of the anisotropic LF
factor.}
\end{table*}

The resonant wavelength of the LF factor, $\lambda_{res}$,
corresponds to setting the real part of the denominator in Eq. 1 to
zero. The resonant increase of the LF factor at ${\omega}$ or
${2\omega}$ results in a many-fold increase of the nonlinear-optical
response of a nanoparticle array.

In this Letter, MSHG assisted by excitation of local surface
plasmons is studied in Co$_{x}$Ag$_{1-x}$ nanogranular films. For
the Co concentration $x<0.45$ these films consist of Co
nanoparticles embedded into Ag matrix. It turns out, that the
excitation of local surface plasmons at the SH wavelength in this
nanomagnetic composite material results in a resonant behavior of
the SHG magnetic contrast.

In the phenomenological description of MSHG [1], the second-order
susceptibility of a magnetic material is considered as a combination
of crystallographic (nonmagnetic) and magnetic terms, which possess
\textit{even} and \textit{odd} parities in magnetization, $M$,
respectively: $\chi_{ijk}^{(2)}=\chi_{ijk}^{(2)
cryst}+\chi_{ijk}^{(2) odd}(M)$, where
$\chi_{ijk}^{(2)cryst}(M)=\chi_{ijk}^{(2)cryst}(-M)$ is the
crystallographic nonmagnetic susceptibility and
$\chi_{ijk}^{(2)odd}(M)$=$-\chi_{ijk}^{(2) odd}(-M)$ is the magnetic
susceptibility.

Nonmagnetic susceptibility is a conventional tensor whereas the
magnetic susceptibility is a pseudotensor. As a consequence, these
susceptibilities possess different sets of tensor elements for the
materials of the same crystallographic symmetry. These tensor
elements of the corresponding susceptibilities are shown in Table 1
for the in-plane isotropic media.

The intensity of MSHG, which is assisted by local plasmon excitation
at the SH wavelength, is given by:

\begin{equation}
\begin{array}{l}
I_{2\omega}(M)\sim[
\textbf{E}_{2\omega}^{cryst}+\textbf{E}_{2\omega}^{odd}(M)]^{2}
\\\sim [\Sigma_{\alpha} \{ L_{\alpha}^{2\omega}(\chi_{\alpha
jk}^{(2)cryst}+\chi_{\alpha jk}^{(2)odd}(M))
f_{j}^{\omega}E_{j}^{\omega}f_{k}^{\omega}E_{k}^{\omega}+c.c.\}]^{2},\label{2}
\end{array}
\end{equation}
where $E_{i}^{\omega}$ is the $i$th component of the fundamental
field, $\textbf{E}_{2\omega}^{cryst}$ and
$\textbf{E}_{2\omega}^{odd}(M)$ are the SH fields originating from
the crystallographic and magnetic susceptibility, respectively, and
$f_{j}^{\omega}$, $f_{k}^{\omega}$ are coefficients that contain
Fresnel factors and linear magneto-optical rotation of polarization
of the fundamental field. The first subscript of tensor elements
$\chi_{ijk}^{(2)}$ is denoted as $\alpha=\perp,\parallel$,
attributing to the corresponding components of the anisotropic LF
factor at the SH wavelength.

Magnetization-induced effects in the SHG are described by the
magnetic contrast:

\begin{equation}
\rho=\frac{I_{2\omega}(M\uparrow)-I_{2\omega}(M\downarrow)}
{I_{2\omega}(M\uparrow)+I_{2\omega}(M\downarrow)}, \label{4}
\end{equation}

where $I_{2\omega}(M\uparrow)$ and $I_{2\omega}(M\downarrow)$are the
values of the SHG intensity for the opposite directions of the
magnetization.

The samples of magnetic nanogranular Co$_{x}$Ag$_{1-x}$ films are
prepared by the co-evaporation of Co and Ag from two independent
electron-beam sources onto glass-ceramic substrates. The structure
of Co$_{x}$Ag$_{1-x}$ films is characterized by X-ray diffraction
and reveals the existence of nanogranules with the diameter ranging
from 3 nm to 6 nm for the composition $\emph{x}<0.4$. The fabricated
granular films exhibit the GMR effect up to 16\%  at room
temperature [13].

The output of an optical-parametric-oscillator (OPO) system
"Spectra-Physics ÌÎÐÎ 710" with tuning range of wavelength from 730
nm to 1000 nm, pulse duration of 4 ns, pulse intensity of 2
MW/cm$^2$, and a Q-switched YAG:Nd$^{3+}$ laser at 1064 nm
wavelength, pulse duration of 15 ns and pulse intensity of 1
MW/cm$^2$ are used as the fundamental radiation. The SHG radiation
reflected from the sample is filtered out by appropriate glass
bandpass and interference filters and is detected by a
photo-multiplier tube and gated electronics. To normalize the SHG
signal over the OPO and YAG:Nd$^{3+}$ laser fluency and the spectral
sensitivity of the optical detection system, an independent
reference arm is used with a Z-cut quartz plate as a reference and a
detection system identical to that in the "sample" arm.

An in-plane dc-magnetic field up to 2 kOe provided by permanent
Fe-Nd-B magnets is applied to the samples in nonlinear
magneto-optical measurements.

The relative phase of $\textbf{E}_{2\omega}^{odd}(M)$ and
$\textbf{E}_{2\omega}^{cryst}$ is measured  by the nonlinear optical
interferometry method described in details elsewhere [14]. The
experimental scheme of the nonlinear optical interferometry is shown
in the inset of Figure 2b. The SH fields from the sample and the
reference interfere at the photomultiplier while the reference SHG
source is translated along the direction of the fundamental beam.
The SHG interferometry is performed by translating a 30 nm thick
indium-tin-oxide (ITO) film on a glass substrate. The interference
pattern, i.e. an oscillating dependence of the SHG intensity as a
function of the reference displacement, results from the phase shift
between the interfering SH fields from the sample and the reference.
Magnetization-induced shift $\theta (M)$ between interference
patterns, which appears due to the reversal of M, is shown on
vectorial diagram in the bottom inset of Figure 1a. This diagram
also shows a phase shift $\Psi (M)$ between
$\textbf{E}_{2\omega}^{odd}(M)$ and $\textbf{E}_{2\omega}^{cryst}$.

\begin{figure}[!h] \vspace{1 cm}
\begin{centering}
\includegraphics
[width=9.5cm]{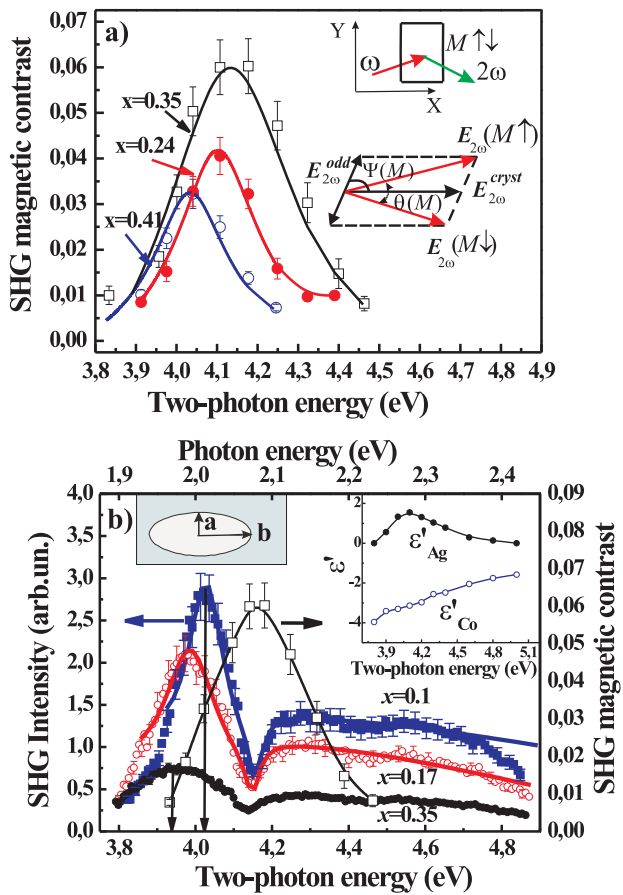} \caption{a) Spectral dependencies of the
SHG magnetic contrast for Co$_{x}$Ag$_{1-x}$ films with
$x$=0.24,0.35,0.41; top inset: geometry of transversal NOMOKE;
bottom inset: vectorial diagram for SH fields, measured for
opposite directions of magnetization and corresponding magnetic
and nonmagnetic contributions to the SH field; b) spectral
dependencies of the nonmagnetic SHG intensity for
Co$_{x}$Ag$_{1-x}$ films with $x$=0.1,0.17,0.35; left-hand inset:
schematic of nanogranular film; right-hand-inset: spectra of
dielectric constants for Co and Ag (from Ref. [15]).}\label{1}
\end{centering}
\vspace{0.1 cm}
\end{figure}

To study wavelength dependence of $\rho$, spectral dependencies of
the SHG intensity are measured for p-in,p-out combination of
polarizations of the fundamental and SH waves for two opposite
directions of $M$ in the configuration of the transversal NOMOKE
(see the top inset of Figure 1a). Figure 1a shows the experimental
spectra of $\rho$ in Co$_{x}$Ag$_{1-x}$ films with $x$=0.24, 0.35,
0.41. All spectra of the SHG magnetic contrast demonstrate
pronounced peaks in the spectral range of the two-photon energy from
3.8 eV to 4.5 eV.

For the interpretation of spectral dependencies of the SHG magnetic
contrast, nonlinear magneto-optical spectroscopy is accompanied by
spectroscopy of the nonmagnetic SHG. Figure 1b shows the set of
spectra of the SHG intensity for nonmagnetized Co$_{x}$Ag$_{1-x}$
films with $x$=0.1, 0.17, 0.35. These spectra of the nonmagnetic SHG
demonstrate a complicated resonant structure: strong and sharp peak
is centered in the vicinity of 4 eV and less intensive broad band
covers the spectral range from 4.2 eV to 4.8 eV.

Resonances in  the SHG intensity and the resonance of the SHG
magnetic contrast demonstrate significant enhancement (by more than
one order of magnitude) of both nonmagnetic SHG and the relative
magnetic contribution. Inset in Figure 1b shows optical spectra of
bulk Co in corresponding spectral range that reveal monotonic
feature-less behavior of the dielectric constant [15]. The lack of
resonant features in optical spectra of Co implies that resonances
of the SHG magnetic contrast and the SHG intensity are not
associated with the bulk magneto-optical and nonlinear optical
properties of Co. Optical properties of metal nanoparticles embedded
to matrix with dielectric constant $\varepsilon>0$ exhibit
additional resonances related to LSPs [16]. Inset in Figure 1b shows
that dielectric constant of Ag is positive in considered spectral
range of two-photon energy from 3.8 eV to 4.5 eV [15]. This implies
that in the aforementioned spectral range one can expect the
appearance of LSP features in optical spectra. Particularly these
LSP modes can be responsible for the resonances of
$L_{\alpha}^{2\omega}$ and corresponding resonances in the spectra
of the nonmagnetic SHG intensity shown in Figure 1b.

We attribute these resonances to the excitation the LSP modes in Co
nanogranules at the SH wavelength. The split of the LSP modes in the
spectral doublet is caused by the deviation of the shape of Co
nanoparticles from spheres [17]. Relation of the resonant features
of the SHG spectra to the LSP modes verifies also by the dependence
of the resonant wavelengths $\lambda_{res}$ on Co concentration.
Increase of $x$ results in the increase of the interparticle
dipole-dipole interaction and red-shift of the LSP resonant
wavelengths which is apparently seen in the set of the SHG spectra
in Figure 1b.

Comparison of the magnetic and nonmagnetic spectra shows that the
resonant wavelength of the SHG magnetic contrast in Figure 1a is
close to the wavelength of the split between  the LSP resonances at
nonmagnetic SHG spectra in Figure 1b. As both crystallographic and
magnetic susceptibilities of bulk Co are not supposed to possess a
resonant behavior in the considered spectral range, the resonance in
the magnetic contrast spectra can be explained by the contribution
to the nonlinear optical response from resonances of the LF factor
corresponding to the LSP modes.

The models, which are considered for the LF factor and NOMOKE with
the LSP assistance, are used for the approximation of the
experimental spectroscopic results.
\begin{figure}[!h] \vspace{1 cm}
\begin{centering}
\includegraphics
[width=9.5cm]{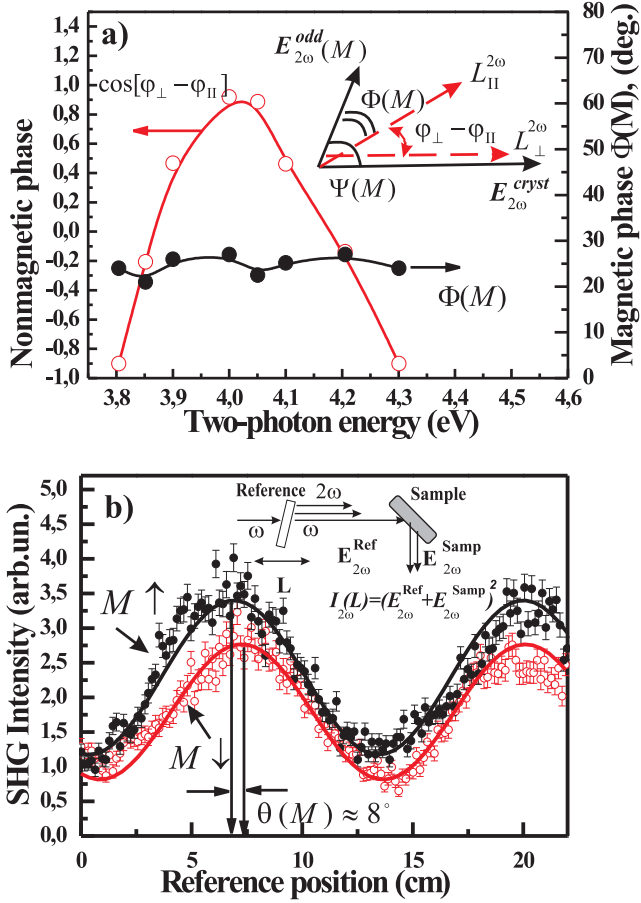} \caption{a) Spectral dependencies of
nonmagnetic and magnetic phases calculated for
Co$_{0.35}$Ag$_{0.65}$ film; inset: vectorial diagram for magnetic
and nonmagnetic SH fields and anisotropic local field factors; b)
Row SHG interference patterns measured for opposite directions of
magnetization for Co$_{0.35}$Ag$_{0.65}$ film; inset: schematic
view of nonlinear optical interferometer.}\label{1}
\end{centering}
\vspace{0.1 cm}
\end{figure}
Solid lines in Figure 1b are the result of the approximation of the
nonmagnetic SHG spectra with Eqs. 1 and 2, that take into account
spectral dependence of $L_{\alpha}^{2\omega}$. Approximation of the
set of spectra for variations of Co content shows that spectral
features of nonmagnetic SHG can be attributed to the resonances of
$L_{\alpha}^{2\omega}$. Two trends in the nonmagnetic SHG spectra
red-shift of the resonant wavelength and the decrease of the
resonant amplitude for the decrease of the Co content, correspond to
the conclusions of the LSP assistance due to the local field
factors. The major adjusting parameter of the approximation is the
ratio of spheroid semiaxes $a/b$. The complete set of nonmagnetic
SHG spectra with variations of $x$ from 0.1 to 0.41 is well
approximated with the same value of $a/b = 0.5 \pm 0.1$, that is a
strong argument for the correctness of the plasmon model in the case
of Co$_{x}$Ag$_{1-x}$ granular films.

For the explanation of the spectral behavior of the SHG magnetic
contrast, a simple model is developed which takes into account the
assistance of NOMOKE in granular films by local surface plasmons in
magnetic nanoparticles. In this model we consider Co$_{x}$Ag$_{1-x}$
granular film as a three-dimensional array of small Co spheroids
with semiaxes $a$ and $b$ parallel and perpendicular to the surface,
respectively, embedded in Ag matrix (see the left-hand inset in
Figure 1b). We suppose that a plasmon resonance is achieved at the
SH wavelength and complex anisotropic LF factors are
$L_{\alpha}^{2\omega}=C_{\alpha} \exp^{-i\varphi_{\alpha}}$,where
$C_{\alpha}$ and $\varphi_{\alpha}$ are real amplitude and phase of
the anisotropic LF factor, respectively. From the symmetry analysis
summarized in Table 1 one can suppose that the predominant first
subscript for the elements of the magnetic susceptibility of the
in-plane isotropic film is $\alpha=\parallel$ whereas $\alpha=\perp$
is predominant as a first subscript for the elements of the
nonmagnetic susceptibility. Together with the assumption that
$\chi_{\perp jk}^{(2)even}\gg\chi_{\parallel jk}^{(2)odd}(M)$, the
MSHG magnetic contrast for the case of the LSP assistance is given
by:

\begin{equation}
\displaystyle \rho\sim  \sum_{j,k} \frac{\chi_{\parallel
jk}^{(2)odd}(M)C_{\parallel}}{\chi_{\perp jk}^{(2)even}C_{\perp}}
\times
\\\cos[\varphi_{\perp}-\varphi_{\parallel}+\Phi(M)],
\label{3}
\end{equation}

where $\Phi(M)$ is the relative phase between
$\textbf{E}_{2\omega}^{odd}(M)$ and complex LF factor
$L_{\parallel}^{2\omega}$ as this is shown on the vectorial diagram
in the inset of Figure 2a. The first subscripts of tensor elements
$\chi_{ijk}^{(2)}$ are denoted as $\parallel$ and $\perp$ to be
attributed to corresponding components of the anisotropic LF factor.

Using this model, the mechanism of the spectral dependence of the
SHG magnetic contrast can be described. In fact, the model
approximation of nonmagnetic SHG spectra gives numerical results for
the spectral dependencies of the amplitudes and relative phase of
local field factors,
$\Delta\varphi=(\varphi_{\perp}-\varphi_{\parallel})$, in the
spectral range of the LSP modes (see vectorial diagram at inset in
Figure 2a). In turn, these data together with experimental data on
the spectral dependence of the SHG magnetic contrast allow one to
calculate, on the base of Eq. 4, spectral dependencies of the
magnetic, $\Phi(M)$, and nonmagnetic,
$\Delta\varphi=(\varphi_{\perp}-\varphi_{\parallel})$, phases.
Figure 2a shows these spectra for Co$_{0.35}$Ag$_{0.65}$ film that
demonstrate resonance spectral dependence of the nonmagnetic phase
and spectral independent behavior of the magnetic phase. Spectral
dependence of the nonmagnetic phase shows a good qualitative
agreement with the spectrum of the SHG magnetic contrast. Moreover,
for off-resonance conditions, the magnetic phases $\Phi(M)$ and
$\Psi(M)$ should coincide, which can be checked by the SHG
interferometry. Figure 2b shows the row interference patterns for
the opposite directions of M in Co$_{0.35}$Ag$_{0.65}$ film that
demonstrate a clear magnetic shift of the phase $\theta (M)=
8^{\circ}$ of SH field. The value of $\Psi(M)= 40^{\circ}\pm
10^{\circ}$ extracted from the shift of interferometric patterns is
in satisfactory agreement with results of calculation for the
resonant spectral range in Figure 2a. Thus, the qualitative
agreement between spectral dependencies of $\rho$  and
$\Delta\varphi$ and approximate equality of $\Phi(M)$ and $\Psi(M)$
allow one to associate the resonances in nonlinear magneto-optical
response of Co$_{x}$Ag$_{1-x}$ films with the assistance from LSP
modes exited in magnetic nanogranules.

In conclusion, local surface plasmons are observed in magnetic
all-metal Co$_{x}$Ag$_{1-x}$ nanogranular films. The excitation of
local surface plasmons in magnetic Co nanoparticles mediates
magnetization-induced SHG resulting in a strong resonance increase
of both the MSHG magnetic contrast and the SHG intensity. This
enhancement of MSHG in magnetic nanoparticles is due to the
assistance of the nonlinear optical effect from the LSP modes and of
a coherent (phase-dependent) interference of magnetic and
nonmagnetic contributions to the SH fields.

$^{\ast \ast}$Electronic address: aktsip@shg.ru; URL:
http://www.shg.ru.

\end{document}